\documentstyle[aps,prl,preprint,floats,epsfig]{revtex}

\begin{document}

\preprint{\tighten\vbox{\hbox{\hfil CLNS 01/1759}
                        \hbox{\hfil CLEO 01-19}
}}

\title{Measurement of the Masses and Widths of the $\Sigma_c^{++}$ and
$\Sigma_c^{0}$ Charmed Baryons}

\author{CLEO Collaboration}
\date{October 29, 2001}

\maketitle
\tighten

\begin{abstract} 
Using data recorded by the CLEO II and CLEO II.V detector configurations at CESR, we
report new measurements of the masses of the $\Sigma_c^{++}$ and $\Sigma_c^0$
charmed baryons, and the first measurements of their intrinsic widths.
We find
$M(\Sigma_c^{++})-M(\Lambda_c^+)=167.4\pm0.1\pm0.2\ {\rm MeV}$,
$\Gamma(\Sigma_c^{++})=2.3\pm0.2\pm0.3\ {\rm MeV}$, and
$M(\Sigma_c^{0})-M(\Lambda_c^+)=167.2\pm0.1\pm0.2\ {\rm MeV}$,
$\Gamma(\Sigma_c^{0})=2.5\pm0.2\pm0.3\ {\rm MeV}$, where the uncertainties
are statistical and systematic, respectively.

\end{abstract}
\newpage

{
\renewcommand{\thefootnote}{\fnsymbol{footnote}}


\begin{center}
M.~Artuso,$^{1}$ C.~Boulahouache,$^{1}$ K.~Bukin,$^{1}$
E.~Dambasuren,$^{1}$ R.~Mountain,$^{1}$ T.~Skwarnicki,$^{1}$
S.~Stone,$^{1}$ J.C.~Wang,$^{1}$
A.~H.~Mahmood,$^{2}$
S.~E.~Csorna,$^{3}$
% ...
\end{center}
 
\small
\begin{center}
$^{1}${Syracuse University, Syracuse, New York 13244}\\
$^{2}${University of Texas - Pan American, Edinburg, Texas 78539}\\
$^{3}${Vanderbilt University, Nashville, Tennessee 37235}\\
% ...
\end{center}

\setcounter{footnote}{0}
}
\newpage


In recent years there have been great advances in charmed baryon spectroscopy.
However, the spin and parity of none of the states has been directly measured,
and we rely upon the pattern of masses of the detected particles, together with their
decay properties, to identify the different states.
The existence of the $J^P={1\over{2}}^+$ $\Sigma_c$ states, which can be considered as
a spin-1 light diquark in combination with a charmed quark, is now
well established.
In 1996, CLEO published\cite{SCS} measurements of the masses and widths of the
analogous $J^P={3\over{2}}^+$ states, the $\Sigma_c^{*++}$ and $\Sigma_c^{*0}$. In
the Heavy Quark Symmetry\cite{ISGUR} picture of heavy hadrons, the decays of the
$\Sigma_c^*$ and $\Sigma_c$ states are closely analogous, and differ in rate
only by calculable phase space and
numerical factors. Previous studies\cite{FOCUS,PDG} of the $\Sigma_c$ baryons have not had 
sufficient
detector resolution to measure their intrinsic widths. In this Letter, using a large sample
of $\Lambda_c^+$ candidates found using the CLEO detector, we are now able to
measure the shape of the $\Sigma_c^{++}$ and $\Sigma_c^{0}$ baryons using the mass
differences
$M(\Lambda_c^+\pi^{\pm}) - M(\Lambda_c^+)$, and extract values of
$\Gamma(\Sigma_c^0)$ and $\Gamma(\Sigma_c^{++})$.

The data presented here
were taken by the CLEO II and CLEO II.V detectors operating at the Cornell
Electron Storage Ring.
The sample used in this analysis corresponds to
an integrated luminosity of 13.7 fb$^{-1}$
taken on the $\Upsilon(4S)$
resonance and in the continuum at energies just below
the $\Upsilon(4S)$.
Of this data, 4.7 fb$^{-1}$ was taken with the CLEO II detector, which is
described in detail elsewhere\cite{KUB}.
We detect charged tracks with a cylindrical drift chamber system inside
a solenoidal magnet, and we detect photons using an electromagnetic
calorimeter consisting of 7800 cesium iodide crystals.
The remainder of the data were taken with the CLEO II.V detector\cite{HILL} 
which is an incremental upgrade of CLEO II, and
incorporates a high resolution silicon vertex detector inside the CLEO II drift
chamber system.

In order to obtain large statistics, we reconstructed $\Lambda_c^+$
baryons using 15 different decay modes\footnote
{Charge conjugate modes are implicit throughout.}.
Measurements of the relative branching
fractions into these modes have previously been presented by the CLEO
collaboration\cite{LAMC}, and the general procedures for finding
those decay modes can be found in these references.
For this search and data set, the exact analysis used has been optimized
for high efficiency and low background.
Briefly, particle identification of $p,K^-$, and $\pi^{+}$ candidates was performed
using specific ionization measurements in the drift chamber,
and when present, time-of-flight measurements. Hyperons and $K^0_S \to \pi^+\pi^-$
decays were found by
detecting their decay points separated from the main event vertex. Photons were
detected using the
CsI electromagnetic calorimeter.

We reduce the combinatorial background, which is highest for
$\Lambda_c^+$ candidates with low momentum, by applying a cut on
$x_p$, where $x_p=p/p_{max}$, $p$ is the momentum
of the charmed baryon, $p_{max}=\sqrt{E^2_{beam}-M^2},$ $M$ is the mass of the
$\Lambda_c^+$ candidate, and $E_{beam}$ is the
beam energy. Using a cut of $x_p > 0.5$
(charmed baryons produced from decays of $B$ mesons are
kinematically limited to $x_p < 0.4$),
we fit each of the invariant mass distributions for these modes to a sum
of a Gaussian signal and a low-order polynomial background.
Combinations within $1.6$ standard deviations of the observed $\Lambda_c^+$ mass peak
are taken as $\Lambda_c^+$ candidates,
where the resolution of each decay mode is taken from a
Monte Carlo simulation
(for the two data sets separately), and
the $\Lambda_c^+$ candidates were kinematically constrained to the
$\Lambda_c^+$ peak mass. 
In this $x_p$ region,
we find a total $\Lambda_c^+$ yield of
$ 58,300\pm380$, and a signal to background ratio of
approximately $1:1.2$. This sample of $\Lambda_c^+$ decays is the same as used
in our analysis of the $\Sigma_c^+$ and $\Sigma_c^{*+}$\cite{SCS2}.
The $x_p$ cut described above was used only to obtain these measures of the
$\Lambda_c^+$ sample, and
was released before continuing with the analysis as we prefer to apply
an $x_p$ cut only on the $\Lambda_c^+\pi^{\pm}$ combinations.

The $\Lambda_c^+$ candidates  were then
combined with each remaining charged $\pi$ track in the event
and the mass difference
$\Delta M = M(\Lambda_c^+\pi^{\pm})-M(\Lambda_c^+)$ was calculated.
To optimize the resolution in this quantity, we calculated an event-by-event
vertex point with those well measured tracks in the event which were
consistent with coming from the beamspot, and then refit the $\Lambda_c^+$ and
$\pi^{\pm}$ trajectories to come from this point. The main effect of this
procedure was to improve the polar angle resolution of the $\pi^{\pm}$, and thus
improve the mass-difference resolution.
Those combinations that were inconsistent with coming from this point were
rejected.
We placed an $x_p > 0.5$ cut on the $\Lambda_c^+\pi^{\pm}$
combination.

Both of the mass-difference spectra (Fig. 1), show a clear
peak of about 2000 events
around 167 $ \rm{MeV}$ due to $\Sigma_c\to\Lambda_c^+\pi^{\pm}$ decays.
These distributions were each then fit to the sum of a
polynomial background with a threshold suppression and a p-wave Breit-Wigner
function
convoluted with a double-Gaussian detector resolution function.
We use a formalism of the Breit-Wigner signal function with
a mass-dependent width, $\Gamma(M) \propto {\Gamma_0 {( {P\over{P_0}})}^3}$, where
$P$ is the $\pi$ momentum in the $\Sigma_c$ rest frame, and $P_0$ is the
$\pi$ momentum calculated at the pole mass; we have tried relativistic and non-relativistic
formalisms of the function and found negligible differences in our results.

The
parameters of the two Gaussians in the resolution function were
$\sigma_1=0.461$ MeV, $\sigma_2= 1.35 $ MeV, and $area_2/area_1 = 0.31$.
These parameters were
found from a Monte Carlo
simulation using a GEANT-based program for the detector response. The
generated Monte Carlo data used the CLEO II and CLEO II.V configurations in the same proportions
as the real data, and assumed a zero-width $\Sigma_c$.
The two Gaussians do not represent the $\Delta M$ resolution of each of the
two different configurations, neither of which has a
resolution which is  well
described by a single Gaussian function.
The solid lines in Figure 1 show the best fits to the data distributions;
the extracted values using these fits are
$M(\Sigma_c^{++})-M(\Lambda_c^+)=167.4\pm0.1\ {\rm MeV}$,
$\Gamma(\Sigma_c^{++})=2.3\pm0.2\ {\rm MeV}$ and
$M(\Sigma_c^{0})-M(\Lambda_c^+)=167.2\pm0.1\ {\rm MeV}$,
$\Gamma(\Sigma_c^{0})=2.5\pm0.2\ {\rm MeV}$.
The dashed lines show the best fits achievable using only the
resolution function to describe the shape of the signal peaks,
and no intrinsic width. The $\chi^2$
of these latter fits are clearly unacceptable.

We have investigated many potential sources of systematic uncertainty in our
measurement of the widths of these particles. We have
analyzed our two data sets independently, using two different double-Gaussian
resolution functions, and find
statistically consistent results. The Monte Carlo studies indicate that the
largest part of the detector
resolution is from the determination of the trajectory of the $\pi^{\pm}$
trajectory rather than the measurement of the $\Lambda_c^+$ daughters;
thus, as expected,
the analysis produces consistent results for different
$\Lambda_c^+$ decay modes.
We assign a 15\% uncertainty in the width of the resolution function,
which translates into a 0.15 MeV uncertainty in the measurement of $\Gamma(\Sigma_c)$.
This is a conservative estimate of the width uncertainty and is based upon the
width measurements of the $\Lambda_{c1}(2625)$ which has
similar kinematics to the $\Sigma_c$.
In order to obtain a width of zero in the data, we would have to use a resolution
function three times wider than that derived from Monte Carlo studies.
We have also fit the mass-difference distributions to resolution functions
varying from a single Gaussian to the
sum of five Gaussians. The extracted $\Sigma_c$ widths from the data
vary by 0.15 MeV when changing from
single to double-Gaussian resolution functions, but are stable with the addition of
further functions. We therefore assign 0.15 MeV
as our uncertainty due to our imperfect knowledge
of the shape of this resolution function.

The polynomial background shape we use is a good fit to the data, but we realize that this
background may include non-phase space contributions arising from feed-down from
other decays of excited charmed baryons, some of which are as yet undiscovered.
For example, some of the $\Sigma_c$ yield is due to
$\Lambda_{c1}^+(2593)\to\Sigma_c\pi$ decays,
and these may have a distorted $\Delta M $ shape
due to the limited phase space available. However, if we place a veto on
decays we observe to be from
this source, our result changes by less than 0.15 MeV.
We have also investigated a veto
of those $\Lambda_c^+\pi^{\pm}$ combinations that are consistent
with being due to $\Lambda_{c1}(2630)$
decays, with a similar null result.
We have performed a large number of fits to the
data with different background parameterizations, as well as different requirements on, for
instance, the $x_p$ of the combinations and the momentum of the $\pi^{\pm}$, and note
only small
variations in the extracted width of the $\Sigma_c$.
From all these studies, we estimate the
systematic uncertainty to be $\pm0.2$ MeV from uncertainties of the effect of feed-down from
other particles,
and a total systematic uncertainty of $\pm0.3$ MeV from all sources.

The measurements of the mass difference, $M(\Sigma_c)-M(\Lambda_c^+)$,
are stable to changes in the background shape and the signal
resolution function. A change from using a Gaussian signal function
(as previous analyses have done),
to a Breit-Wigner function, produces a shift of only 0.02 MeV in the extracted pole mass.
Overall, including all systematic uncertainties in the fitting procedure,
feed-down effects
from the $\Lambda_{c1}(2593)$ and momentum measurements,
we estimate a total systematic uncertainty of $\pm 0.2$  MeV on the
measured mass differences. Much of this systematic uncertainty cancels in the
comparison of the two mass differences, giving an isospin mass
splitting of $M(\Sigma_c^{++})-M(\Sigma_c^0)=0.2\pm0.1\pm0.1\ $ MeV. This result is
consistent both with the recent measurement by FOCUS\cite{FOCUS}, as well
as earlier results\cite{PDG}
that all indicate a small isospin splitting between these states.
Theoretical models
predict values of this mass splitting that vary from -3 to +3 MeV\cite{Theory}.

Using scaling laws and measures of the non-charmed $\Sigma$ widths, Rosner\cite{ROS} has
predicted a value for the widths of the $\Sigma_c^{++}$ and
$\Sigma_c^0$ of $1.3 {\ \rm MeV}$, and Huang {\it et al.} \cite{HUANG}
have predicted widths of around $2.4 {\ \rm MeV}$.
Tawfiq and collaborators \cite{TAWFIQ} use strange-baryon data and a Light-Front Quark Model
to predict $\Sigma_c$ widths of $1.6 {\ \rm MeV}$, whereas Ivanov {\it et al.}\cite{IVANOV}, use
a Relativistic Three-Quark Model to predict $\Sigma_c$ widths of around $2.7 {\ \rm MeV}$.
Pirjol and Yan\cite{PIRJOL}
have directly scaled from the measured $\Sigma_c^*$ widths as input,
and derived $\Sigma_c$ widths of 2.0 MeV.
Our results are consistent with these predictions, all of which use the Heavy Quark
Symmetry model of baryon structure and decays.

In conclusion, we present new measurements of the masses of the $\Sigma_c^{++}$
and $\Sigma_c^0$ charmed baryons relative to the $\Lambda_c^+$ mass. We measure
$M(\Sigma_c^{++})-M(\Lambda_c^+)=167.4\pm0.1\pm0.2\ {\rm MeV}$ and
$M(\Sigma_c^{0})-M(\Lambda_c^+)=167.2\pm0.1\pm0.2\ {\rm MeV}$.
These measurements of the masses of the $\Sigma_c^{++}$ and $\Sigma_c^0$
are the most statistically
precise available and are consistent with the world average values. They supercede the
previous CLEO II numbers\cite{OLD} which used a subset of the present data set.
We make the first measurements of
the intrinsic widths of these particles, and find
$\Gamma(\Sigma_c^{++})=2.3 \pm 0.2 \pm0.3\ {\rm MeV}$
and $\Gamma(\Sigma_c^0)=2.5 \pm 0.2 \pm0.3\ {\rm MeV}$.
These widths are consistent with theoretical expectations.
\vskip 1 cm
\centerline{\bf ACKNOWLEDGMENTS}
\vskip 0.5cm
We gratefully acknowledge the effort of the CESR staff in providing us with
excellent luminosity and running conditions.
M. Selen thanks the PFF program of the NSF and the Research Corporation, 
and A.H. Mahmood thanks the Texas Advanced Research Program.
This work was supported by the National Science Foundation, and the
U.S. Department of Energy.

\begin{figure}[htb]
\noindent
\psfig{bbllx=0pt,bblly=0pt,bburx=490pt,bbury=490pt,
file=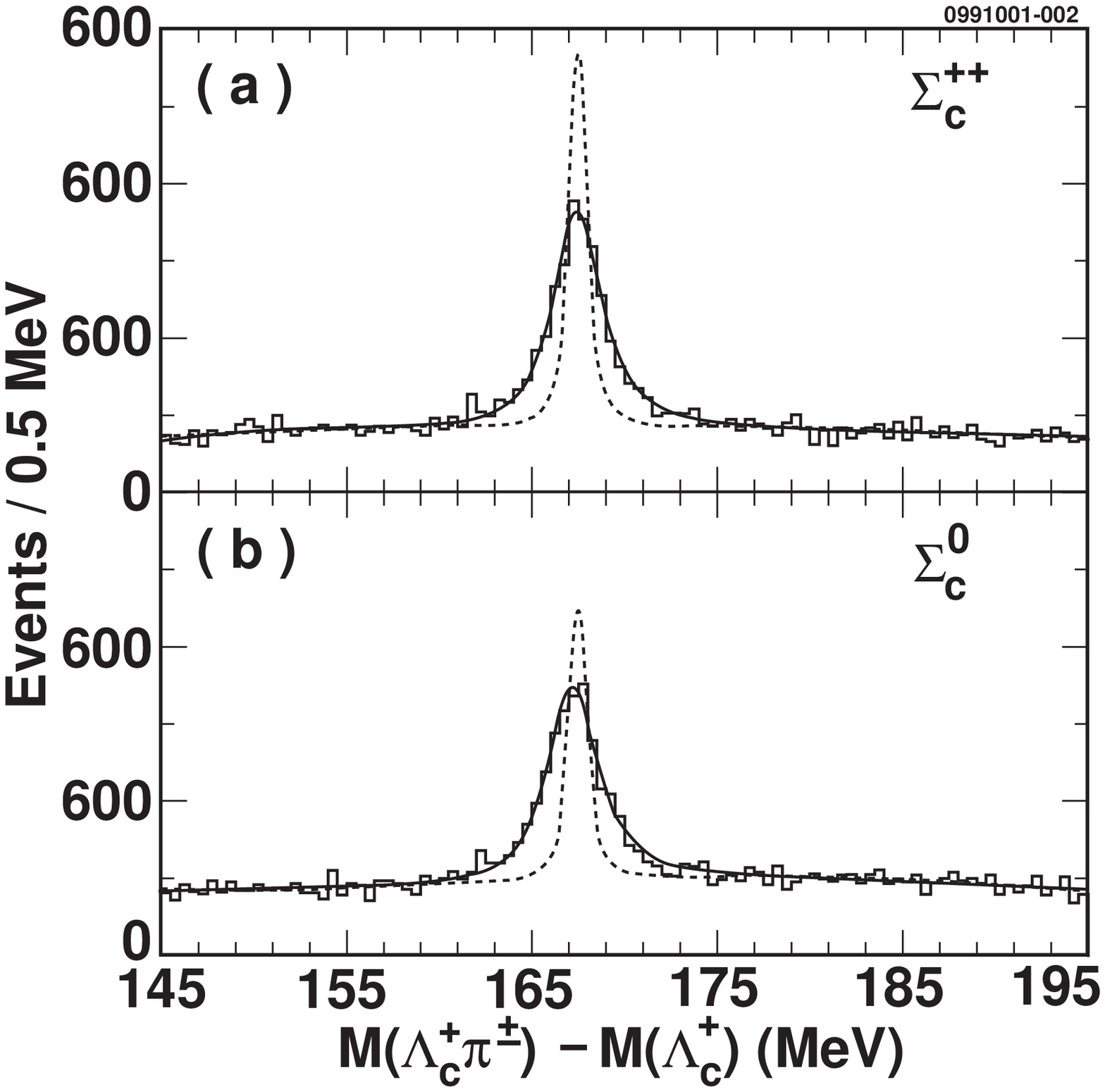,width=6.0in}
\caption[]{Mass-difference spectra for (a) $\Lambda_c^+\pi^+$, and
(b) $\Lambda_c^+\pi^-$.
The lines shown are the results of fits
allowing for a $\Sigma_c$ intrinsic width (solid)
and with no $\Sigma_c$ intrinsic width (dashed).}
\end{figure}
\begin{figure}[htb]
\noindent
\end{figure}


\narrowtext

\smallskip









\begin{references}
\bibitem{SCS} G.~Brandenberg  {\it et al.}, Phys. Rev. Lett. {\bf 78}, 2304 (1997).

\bibitem{ISGUR} N.~Isgur and M.~Wise, Phys. Rev. Lett. {\bf 66}, 1130 (1991).

\bibitem{FOCUS} J.~Link  {\it et al.}, Phys. Lett. {\bf B488}, 218 (2000).

\bibitem{PDG}  D.~Groom  {\it et al.}, Eur. Phys. J. {\bf C15}, 1 (2000).

\bibitem{KUB} Y.~Kubota  {\it et al.}, Nucl. Instr. and Meth. {\bf A320}, 66 (1992).

\bibitem{HILL} T.~Hill  {\it et al.}, Nucl. Instr. and Meth. {\bf A418}, 32 (1998).

\bibitem{LAMC} P.~Avery  {\it et al.}, Phys. Rev. {\bf D43}, 3599 (1991); P.~Avery  {\it et al.},
Phys. Rev. Lett. {\bf 71}, 2391 (1993); P.~Avery  {\it et al.}, Phys. Lett. {\bf B235}, 257 (1994);
M.S.~Alam  {\it et al.}, Phys. Rev. {\bf D57}, 4467 (1998).

\bibitem{SCS2} R.~Ammar  {\it et al.}, Phys. Rev. Lett. {\bf 86}, 1167 (2001).

\bibitem{Theory} S.~Capstick, Phys. Rev. {\bf D36}, 2800 (1987);
L.~H.~Chan, Phys. Rev {\bf D31}, 204 (1985); W.Y.P.~Hwang and D.B.~Lichtenberg, Phys. Rev.
{\bf D35}, 3526 (1987); S.N.~Sinha, S.M.~Sinha, M.~Rahman and D.Y.~Kim,
Phys. Lett. {\bf 218B}, 333 (1999).

\bibitem{ROS} J.~Rosner, Phys. Rev. {\bf D52}, 6461 (1995).

\bibitem{HUANG} M-Q.~Huang, Y-B.~Dai and C-S.~Huang, Phys. Rev. {\bf D52}, 3986 (1995).

\bibitem{TAWFIQ} S.~Tawfiq  {\it et al.}, Phys. Rev. {\bf D58}, 054010 (1998).

\bibitem{IVANOV} M.~A.~Ivanov, J.G. K\"orner, V.E. Lyubovitskij and
A.G. Rusetsky, Phys. Lett. {\bf B442}, 435 (1998). 

\bibitem{PIRJOL} D.~Pirjol and T.~M.~Yan, Phys. Rev. {\bf D56}, 5483 (1997).

\bibitem{OLD} G.~Crawford {\it et al.}, Phys. Rev. Lett. {\bf 71}, 3259 (1993).

\end{references}
\end{document}